\documentclass{PoS}

\usepackage{wrapfig}

\title{Nuclear PDFs from the LHeC perspective}

\ShortTitle{Nuclear PDFs from the LHeC perspective}

\author{\speaker{Hannu Paukkunen} $^\dag$ \\%
        University of Jyv\"askyl\"a and Helsinki Institute of Physics, Finland\\
        E-mail: \email{hannu.paukkunen@jyu.fi}}

\author{{Kari J. Eskola} 
        \thanks{KJE and HP acknowledge the financial support from the Academy of Finland, Project No. 133005.}\\
        University of Jyv\"askyl\"a and Helsinki Institute of Physics, Finland\\
        E-mail: \email{kari.eskola@jyu.fi}}

\author{N\'estor Armesto
           \thanks{The work of NA is supported by European Research Council grant HotLHC ERC-201-StG-279579; by Ministerio de Ciencia e Innovaci\'on of Spain under projects FPA2009-06867-E and FPA2011-22776; by Xunta de Galicia (Conseller\'{\i}a de Educaci\'on and Conseller\'\i a de Innovaci\'on e Industria - Programa Incite); by the Spanish Consolider-Ingenio 2010 Programme CPAN and by FEDER.}\\
        University of Santiago de Compostela,
        Departamento de F\'isica de Part\'iculas and IGFAE, Spain \\
        E-mail: \email{nestor.armesto@usc.es}}

\author{For the LHeC Study Group}

\abstract{We study the prospects for constraining the nuclear parton distribution functions
by small-$x$ deep inelastic scattering. Performing a global fit of nuclear parton distribution
functions including a sample of pseudodata representing expected measurements at the planned
LHeC collider, we demonstrate that the accuracy of
the present nuclear parton distributions could be be improved substantially. We also 
discuss the impact of flavour-tagged data.}

\FullConference{XXI International Workshop on Deep-Inelastic Scattering and Related Subjects\\
                 22-26 April, 2013\\
                 Marseilles, France}

\begin{document}

\vspace{-0.2cm}
\section{Introduction}
\label{sec:Introduction}

\vspace{-0.2cm}
Much of the present knowledge of the parton distribution functions (PDFs) of free nucleons
\cite{Forte:2013wc} comes from the measurements of deeply inelastic scattering (DIS) in lepton-nucleon collisions.
Observables like inclusive \cite{Aaron:2009aa} as well as charm and bottom cross-sections
\cite{Aaron:2009af} not only serve as tight PDF constraints, but also lend support to the necessity of a
consistent theoretical treatment of heavy quarks in 
calculating the DIS observables, see e.g ~\cite{Nadolsky:2009ge}.

The program of extracting the nuclear PDFs (nPDFs) \cite{Eskola:2009uj,Hirai:2007sx,deFlorian:2011fp}
also relies much on the DIS data. However, whereas the HERA data \cite{Aaron:2009aa,Aaron:2009af} for the free proton
reach $x \sim10^{-5}$ in perturbative values of $Q^2$, the DIS-measurements for nuclear
targets are bound to severely higher momentum fractions, $x \gtrsim 10^{-2}$. This situation leaves
the nuclear PDFs badly unconstrained and fit-function dependent at small $x$. 
Thus, additional small-$x$ measurements are desperately called for.
Such data would allow to test the collinear factorization for nuclear
targets --- known to work very well for the presently available data --- in a larger kinematical
domain and study the possible onset of the nuclear-enhanced higher-twist corrections \cite{Mueller:1985wy,Qiu:2003vd}.
Here, we intend to demonstrate how dramatically precision-data from the planned electron-proton/ion collider
LHeC \cite{AbelleiraFernandez:2012cc}, could constrain the nPDFs.

\begin{figure*}[ht]
\begin{minipage}[b]{0.48\linewidth}
\centering
\includegraphics[width=\textwidth]{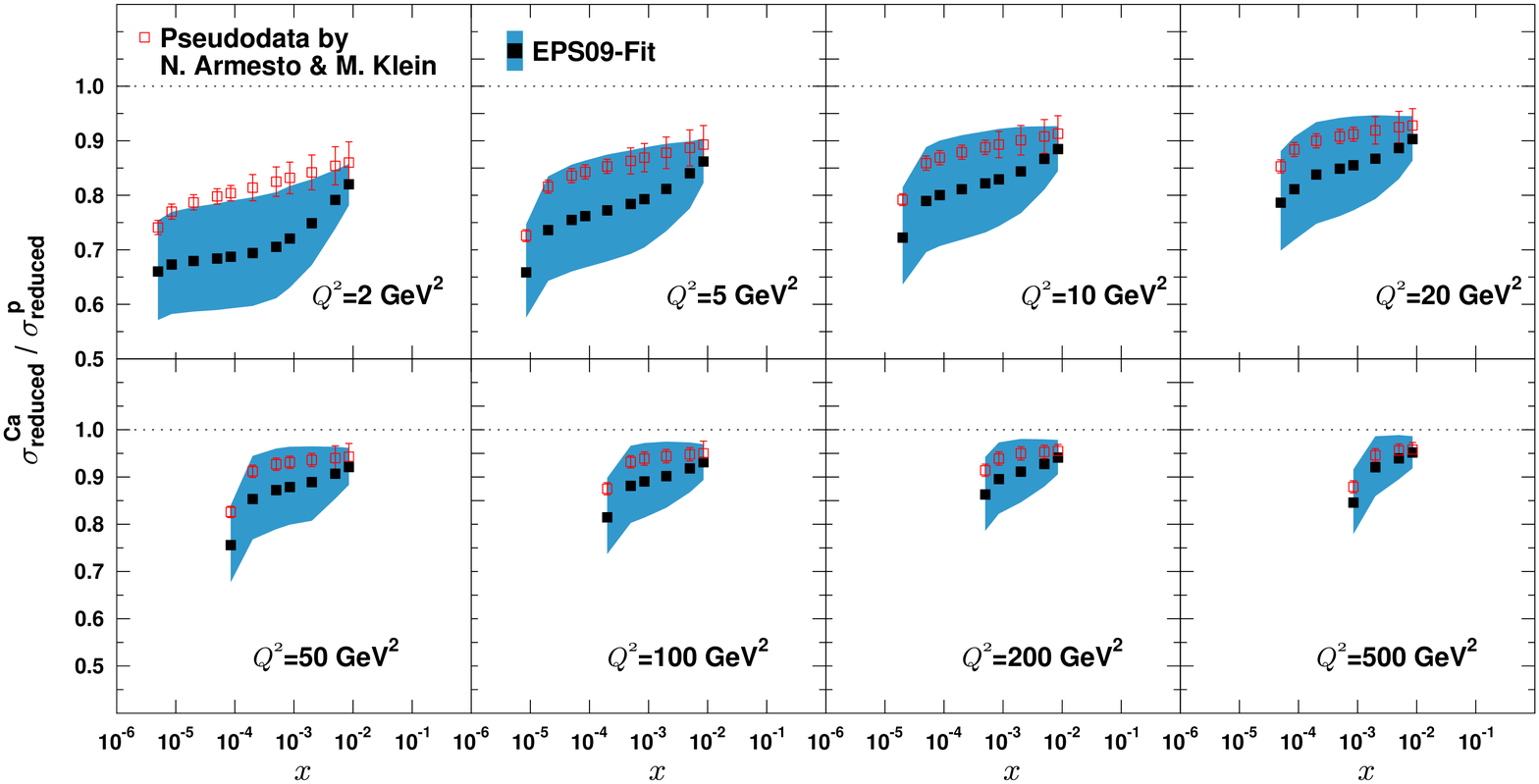}
\end{minipage}
\hspace{1cm}
\begin{minipage}[b]{0.48\linewidth}
\centering
\includegraphics[width=\textwidth]{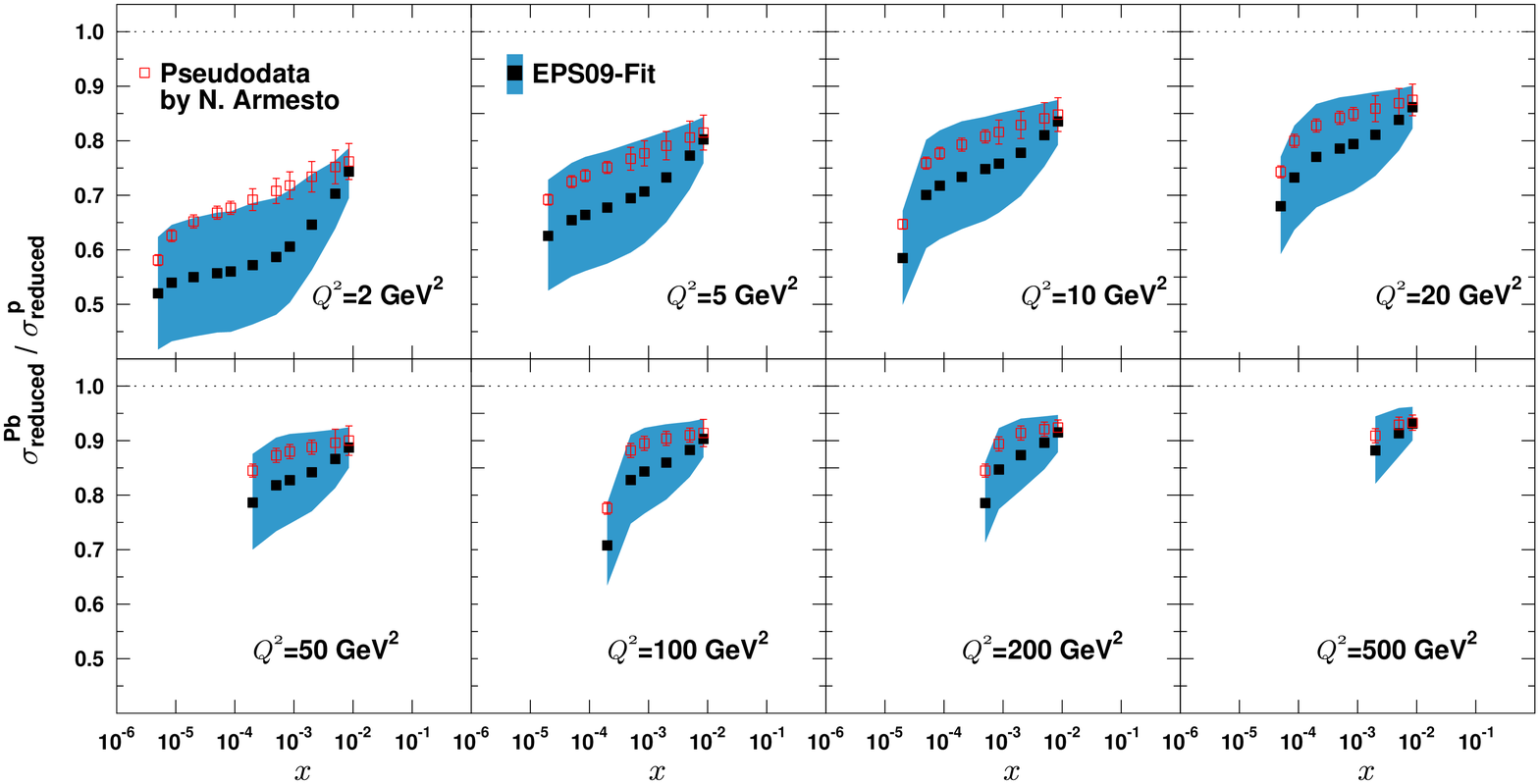}
\end{minipage}
\caption{$\sigma_{\rm reduced}^{\rm Ca} / \sigma_{\rm reduced}^{\rm proton}$ and
$\sigma_{\rm reduced}^{\rm Pb} / \sigma_{\rm reduced}^{\rm proton}$ pseudodata (red open squares) compared to the EPS09-predictions
indicated by the black squares and the blue $\Delta \chi^2=50$ errorbands.
}
\label{Fig:Pseudodata_EPS09}
\end{figure*}

\vspace{-0.2cm}
\section{Fit To Inclusive Pseudodata}
\label{sec:Analysis}

\vspace{-0.2cm}
In order to estimate the effect of high-precision DIS-measurements at the LHeC, 
we employ a sample of pseudodata for the ratios of the reduced, inclusive DIS cross-sections
\begin{equation}
{\sigma_{\rm reduced}^{\rm Ca}(x,Q^2)}/{\sigma_{\rm reduced}^{\rm p}(x,Q^2)}, \quad {\rm and} \quad
{\sigma_{\rm reduced}^{\rm Pb}(x,Q^2)}/{\sigma_{\rm reduced}^{\rm p}(x,Q^2)},
\end{equation}
where
\begin{equation}
\sigma_{\rm reduced}(x,Q^2) = F_2 \left[1 - \frac{y^2}{1+(1-y)^2}\frac{F_L}{F_2} \right],
\label{eq:reduced}
\end{equation}
and $x$, $y$ and $Q^2$ are the standard DIS variables.
Here, the free proton cross-sections ${\sigma_{\rm reduced}^{\rm p}}$ come from a pQCD-based
simulation which was modified for the nuclear case ${\sigma_{\rm reduced}^{A}}$ according to the
model described in \cite{Armesto:2002ny}. The uncertainties include the statistical and
systematic errors estimated to be reachable at the LHeC. We restrict here to
a kinematical region $x<0.01$ and $2 \, {\rm GeV}^2 < Q^2 < 1000 \, {\rm GeV}^2$. These data
are shown in Figure~\ref{Fig:Pseudodata_EPS09}
together with the corresponding predictions from the EPS09 \cite{Eskola:2009uj} nuclear PDFs.
The lepton beam of $50 \, {\rm GeV}$, proton beam of $7000 \, {\rm GeV}$,
Calcium-ions of $3500 \, {\rm GeV}/{\rm nucleon}$, and Lead-ions of  $2750 \, {\rm GeV}/{\rm nucleon}$
have been assumed. The blue bands in these figures represent the EPS09 
uncertainty range. As this band clearly exceeds the pseudodata uncertainties, one
would anticipate that if such data are included in a global analysis, the resulting nPDFs
should be substantially better constrained. At very small $x$, the sudden drop in the ratios is due to the onset of the longitudinal
structure function $F_L$, see Eq.~(\ref{eq:reduced}).

To concretely estimate the level of improvement, we have performed a global fit of nPDFs
similar to EPS09 \cite{Eskola:2009uj}, but including the pseudodata sample introduced above. We refer to
this fit here as the Fit 1. As the model which was 
used to generate these data is in a reasonable agreement with the presently available
DIS data, no severe disagreement occurs between the real and pseudodata included in the
fit. The analysis was carried out in the standard next-to-leading order (NLO) perturbative QCD,
in the variable-flavour-number scheme (SACOT prescription) with CTEQ6.6 \cite{Nadolsky:2008zw} 
set of free proton PDFs as a baseline. In comparison to the EPS09-analysis, one additional
gluon parameter which was freezed in EPS09 was freed, and the only additionally weighted data set was the PHENIX $\pi^0$-data
\cite{Adler:2006wg} to better constrain the large-$x$ gluons.
\begin{wrapfigure}{r}{0.55\textwidth}
\centerline{\includegraphics[width=0.6\textwidth]{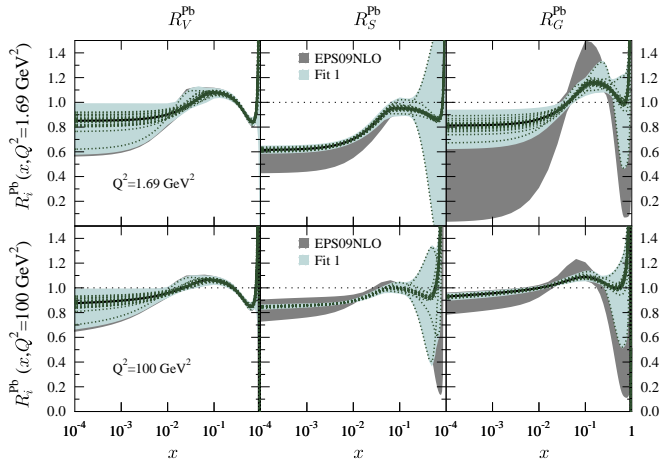}}
\caption[]{\small The nuclear modifications $R_V$, $R_S$, $R_G$ for Lead at $Q^2_0=1.69 \, {\rm GeV}^2$
and at $Q^2=100 \, {\rm GeV}^2$. The thick black lines indicate the best-fit results,
whereas the dotted green curves denote the individual error sets. The shaded blue bands are the $\Delta \chi^2=50$
error bands from  Fit 1, and the grey ones are the EPS09 error bands.}
\label{Fig:PbPDFs}
\end{wrapfigure}

\begin{figure*}[ht]
\begin{minipage}[b]{0.5\linewidth}
\centering
\includegraphics[width=\textwidth]{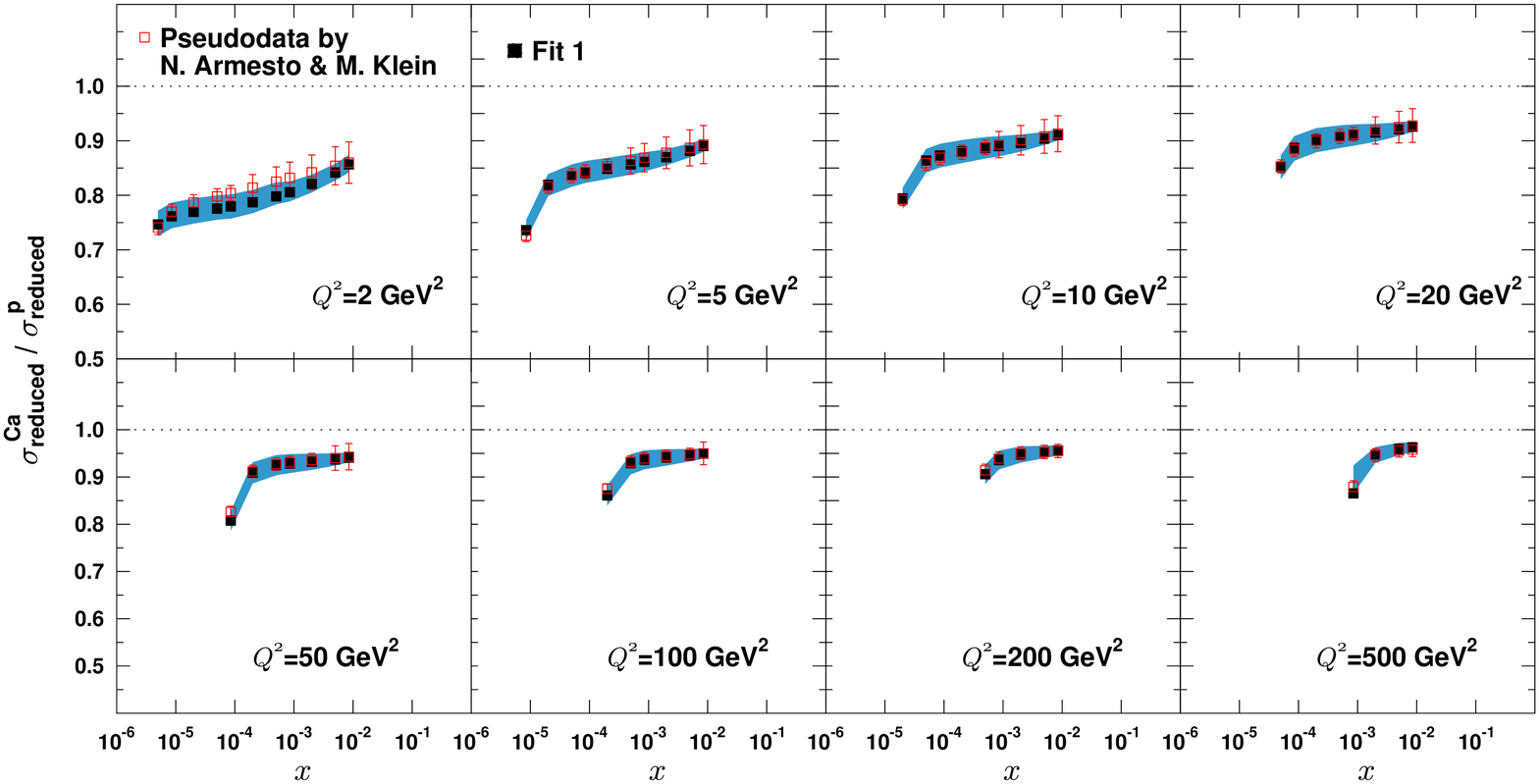}
\end{minipage}
\begin{minipage}[b]{0.5\linewidth}
\centering
\includegraphics[width=\textwidth]{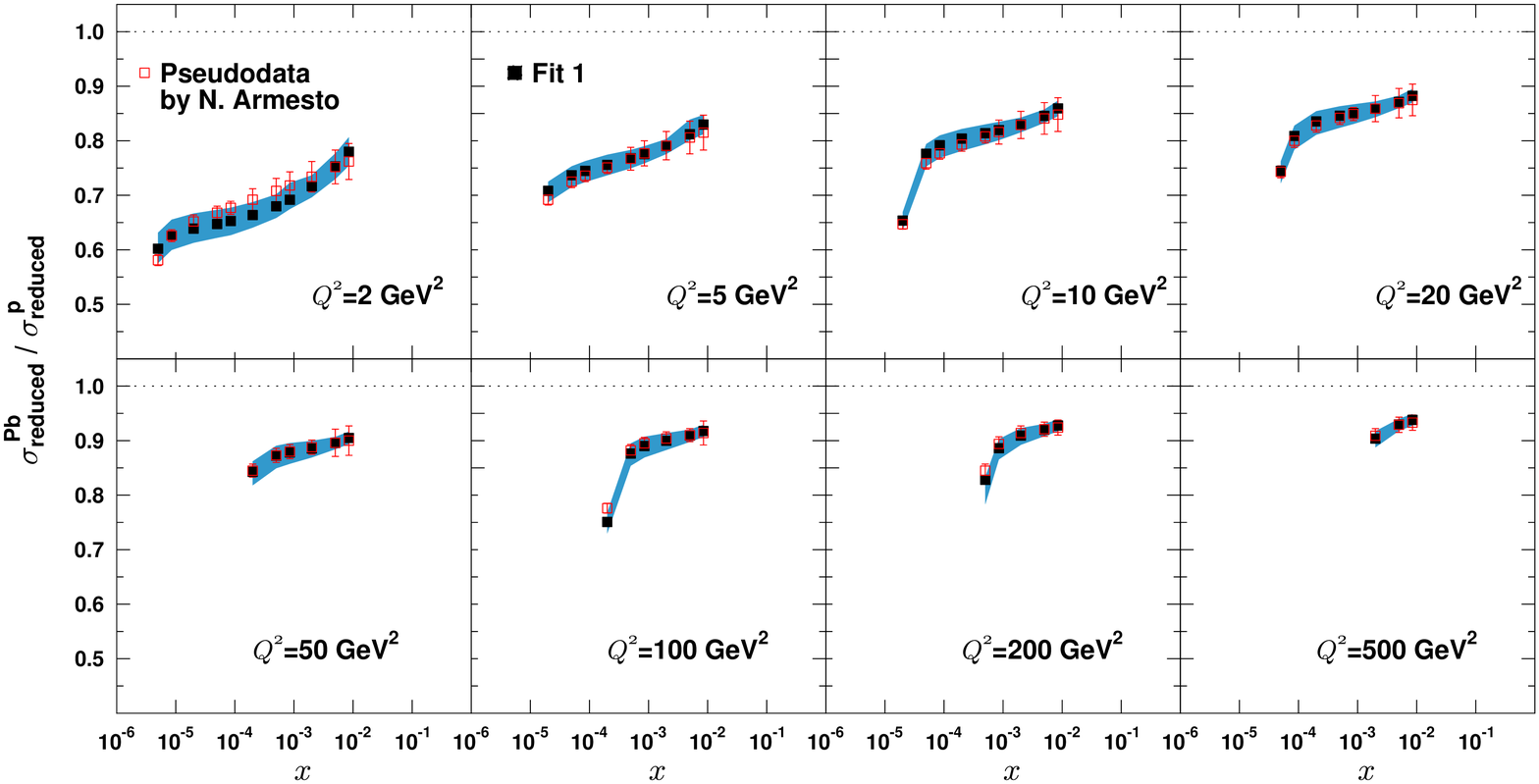}
\end{minipage}
\caption{As Figure~1 but calculated using the nPDFs from  Fit 1.}
\label{Fig:Pseudodata_Fit1}
\end{figure*}

The results of this exercise are depicted in Figures~\ref{Fig:PbPDFs} and \ref{Fig:Pseudodata_Fit1}. First,
in Figure \ref{Fig:PbPDFs}, we show the resulting modifications in the nuclear PDFs $f_{i}^{\rm Pb}(x,Q^2)$
relative to the free proton PDFs $f_{i}^{\rm free \, proton}(x,Q^2)$,
\begin{equation}
R_{i}^{\rm Pb}(x,Q^2) \equiv \frac{f_{i}^{\rm Pb}(x,Q^2)}{f_{i}^{\rm free \, proton}(x,Q^2)},
\label{eq:partondefinition}
\end{equation}
for average valence quarks $R_V^{\rm Pb}$, average sea quarks $R_S^{\rm Pb}$ and gluons $R_G^{\rm Pb}$ at two different scales.
The blue bands correspond to the present analysis, whereas the grey bands are from the EPS09. 
Both bands have been calculated by restricting the $\chi^2$ to remain within $\Delta \chi^2=50$ from
its minimum. Evidently, in the analysis performed here, the small-$x$ uncertainties have been greatly reduced for
both, gluons and sea quarks. While for the quarks this happens mainly because the cross-sections are  directly sensitive
to the quark PDFs, the reduced gluon uncertainty is rather due to the wide $Q^2$-range spanned by the
pseudodata offering an efficient leverage to constrain the gluons through the parton evolution.
In Figure~\ref{Fig:Pseudodata_Fit1} we show how
the Fit 1 compares with the pseudodata and how the substantial EPS09 errorbands
(see Figure~\ref{Fig:Pseudodata_EPS09}) have dramatically shrunk
due to the expected high precision of the data. The small mismatch between the fit
and the pseudodata at the very small $Q^2$ demonstrates that also more flexibility
in the shape of the present form of the fit-functions should be allowed.

\vspace{-0.2cm}
\section{The Flavour Decomposition}
\label{sec:Flavour_Decomposition}

\begin{wrapfigure}{r}{0.6\textwidth}
\vspace{-1.5cm}
\centerline{\includegraphics[width=0.6\textwidth]{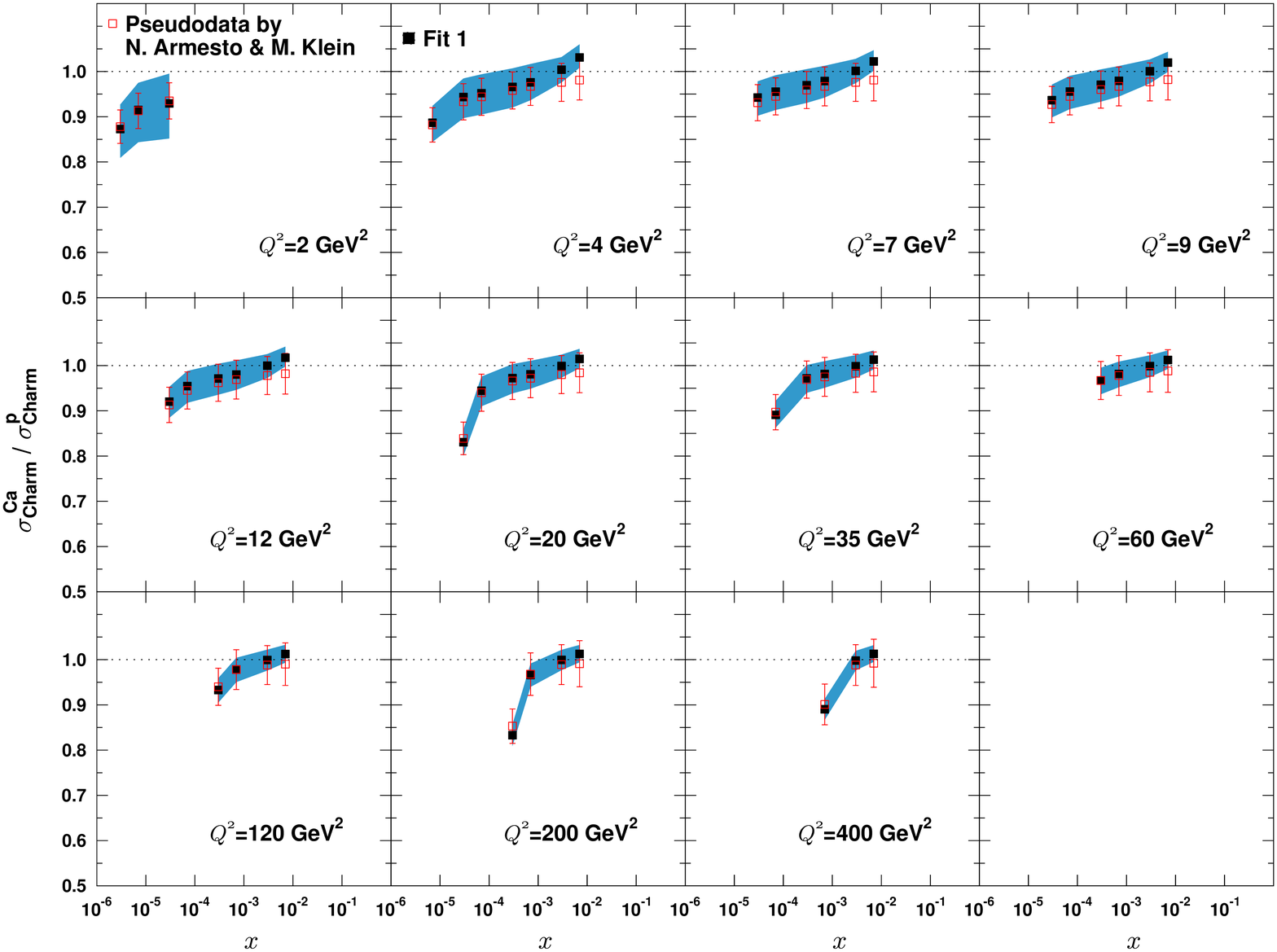}}
\vspace{-0.5cm}
\centerline{\includegraphics[width=0.6\textwidth]{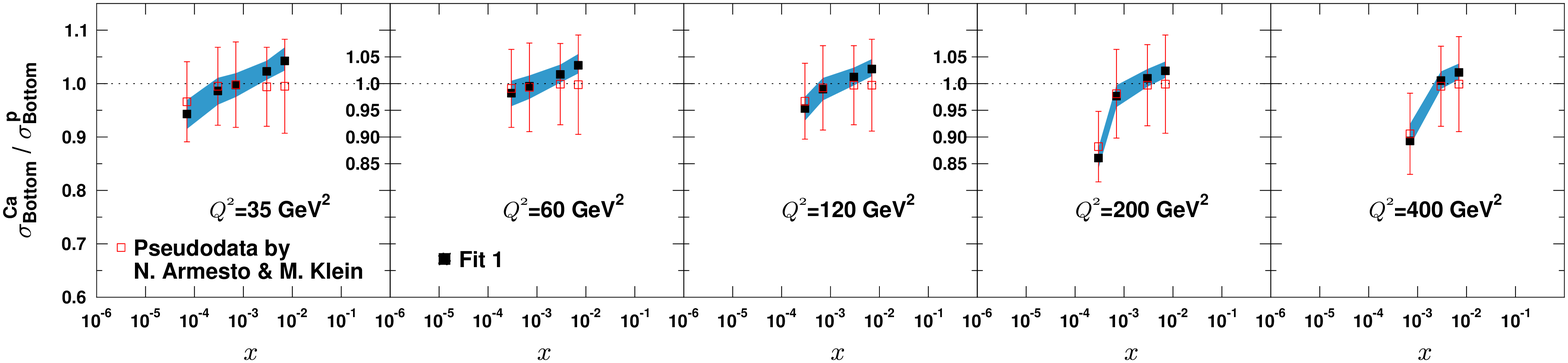}}
\caption[]{\small The Calcium pseudodata (red open squares) for charm- and bottom-tagged
reduced cross-section ratios compared to  Fit 1 predictions (black squares and
blue error bands).}
\label{Fig:Pseudodata_Fit_Ca_Charm_and_Bottom}
\end{wrapfigure}

\begin{wrapfigure}{r}{0.33\textwidth}
\vspace{-1cm}
\centerline{\includegraphics[width=0.35\textwidth]{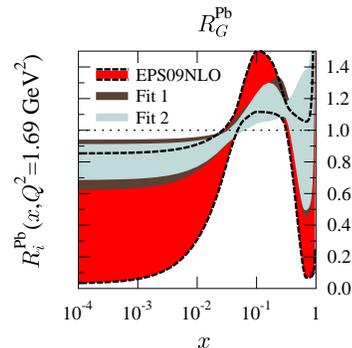}}
\caption[]{\small The nuclear modification $R_G$ for Lead at $Q^2_0=1.69 \, {\rm GeV}^2$ from Fit 1, Fit 2, and EPS09.}
\label{Fig:PbPDFs2}
\end{wrapfigure}

\vspace{-0.2cm}
The experimental identification of the events involving heavy quarks opens further possibilities
 for constraining the nPDFs. Especially, the production of
heavy flavours --- charm and beauty --- are intimately linked to the gluon distributions:
At NLO, the low-$Q^2$ heavy flavours are produced by $g + \gamma^* \rightarrow g + h + \overline{h}$ partonic
process involving the gluon PDFs, but even at high-$Q^2$ the LO subprocess like $h + \gamma^* \rightarrow h + X$
can be ultimately traced back to the gluon PDFs, as it is the gluon splitting that is driving the 
DGLAP evolution of the heavy quark PDFs. Therefore, it is mainly the accuracy of the gluon PDF
that is anticipated to acquire improvement by additional flavour-tagged data. To estimate 
the level of such an improvement we have added a sample of pseudodata for the ratios
\begin{equation}
{\sigma_{\rm reduced, \, charm}^{\rm Ca, \, Pb}(x,Q^2)}/{\sigma_{\rm reduced, \, charm}^{\rm p}(x,Q^2)}
\,\,\,\,\,\,\, {\rm and} \,\,\,\,\,\,\,
{\sigma_{\rm reduced, \, bottom}^{\rm Ca, \, Pb}(x,Q^2)}/{\sigma_{\rm reduced, \, bottom}^{\rm p}(x,Q^2)}
\end{equation}
on top of the inclusive data presented earlier.  The fit to this larger set of data is
called here Fit 2. The foreseen additional constraints from the flavour-tagged 
data can be understood from Figure~\ref{Fig:Pseudodata_Fit_Ca_Charm_and_Bottom}, where
these pseudodata for the Calcium are contrasted with  Fit 1 predictions. Although the expected accuracy
of the flavour-tagged data is less than that of the inclusive data, especially at the
lowest-$Q^2$ panels for charm production there is clearly some room for improvement. 
The size of the resulting improvement after the fit in $R_G^{\rm Pb}$ at low-$Q^2$ 
is shown in Figure~\ref{Fig:PbPDFs2}.

\vspace{-0.2cm}
\section{Conclusion}
\label{sec:Conclusion}

\vspace{-0.2cm}
In short, the present study proves that the realization of an electron-proton/ion
collider like the LHeC would offer a tremendous improvement in pinning down	
the nPDFs. Especially, the determination of the small-$x$ gluons could be
greatly improved by precision DIS data similar to the pseudodata used in this work.
Presently, the  determination of the nuclear gluon PDFs relies on only a very few data sets and
consequently suffers from large uncertainties. For example, the recent EPS09-analysis employs the $\pi^0$ production
data in d+Au collisions \cite{Adler:2006wg} to constrain the nuclear effects in gluon PDFs. This process is,
however, complicated by the presence of the parton-to-pion fragmentation functions which may also
undergo a modification with respect to the free proton fragmentation functions \cite{Sassot:2009sh}
and would affect the outcome of the analysis, as in \cite{deFlorian:2011fp}. In this respect
the pure inclusive DIS provides a much cleaner environment to study the nPDFs.

We note that here we concentrated only on the small-$x$ part of the whole kinematic reach of the LHeC
which would range up to $x\sim 0.6$ and $Q^2 \sim 10^6 \, {\rm GeV}^2$. Also, there are many other nPDF-related
processes that could be measured at the LHeC \cite{AbelleiraFernandez:2012cc} (charged-current reactions, jet production,...).
Therefore, these results should only be taken as the lower limit for the expected impact of the LHeC on the nPDFs.

\end{document}